\documentclass[12pt]{iopart}

\usepackage{hyperref}
\usepackage{graphicx}

\begin{document}
\title[]{Surface oxides, carbides, and impurities on RF superconducting Nb and Nb$_3$Sn: A comprehensive analysis}

\author{Zeming Sun$^{1,2,*}$, Zhaslan Baraissov$^3$, Catherine A. Dukes$^4$, Darrah K. Dare$^5$, Thomas Oseroff$^{1,2}$, Michael O. Thompson$^6$, David A. Muller$^3$, Matthias U. Liepe$^{1,2}$}

\address{$^1$ Cornell Laboratory for Accelerator-based Sciences and Education, Cornell University, Ithaca, New York, 14853, United States of America}
\address{$^2$ Department of Physics, Cornell University, Ithaca, New York, 14853, United States of America}
\address{$^3$ School of Applied \& Engineering Physics, Cornell University, Ithaca, New York, 14853, United States of America}
\address{$^4$ Laboratory for Astrophysics and Surface Physics, University of Virginia, Charlottesville, Virginia, 22904, United States of America}
\address{$^5$ Cornell Center for Materials Research, Cornell University, Ithaca, New York, 14853, United States of America}
\address{$^6$ Materials Science and Engineering, Cornell University, Ithaca, New York, 14853, United States of America}
\ead{zs253@cornell.edu}
\vspace{10pt}
\begin{indented}
\item[]September 2023
\end{indented}

\begin{abstract}
 Surface structures on radio-frequency (RF) superconductors are crucially important in determining their interaction with the RF field. Here we investigate the surface compositions, structural profiles, and valence distributions of oxides, carbides, and impurities on niobium (Nb) and niobium-tin (Nb$_3$Sn) \textit{in situ} under different processing conditions. We establish the underlying mechanisms of vacuum baking and nitrogen processing in Nb and demonstrate that carbide formation induced during high-temperature baking, regardless of gas environment, determines subsequent oxide formation upon air exposure or low-temperature baking, leading to modifications of the electron population profile. Our findings support the combined contribution of surface oxides and second-phase formation to the outcome of ultra-high vacuum baking (oxygen processing) and nitrogen processing. Also, we observe that vapor-diffused Nb$_3$Sn contains thick metastable oxides, while electrochemically synthesized Nb$_3$Sn only has a thin oxide layer. Our findings reveal fundamental mechanisms of baking and processing Nb and Nb$_3$Sn surface structures for high-performance superconducting RF and quantum applications. 
\end{abstract}

%
\vspace{2pc}

\section{Introduction}

Niobium (Nb) oxides with diverse chemistries, phase structures, atomic-scale layering, and properties are widely used in various applications, including catalysis, electrolytic capacitors, solar cells, photodetectors, carrier selective contacts, photochromic and electrical-switching devices, and critical temperature sensors \cite{SunRef1}. The oxides in the form of native layers on the superconducting Nb surface are critical -- functional or detrimental -- in a wide range of emerging superconducting radio-frequency/microwave (SRF) technologies, such as SRF resonators \cite{SunRef12}, SRF cavities and photoinjectors for particle accelerator applications \cite{SunRef3,SunRef6}, superconducting quantum computing \cite{SunRef9,SunRef10}, microwave parametric amplifiers \cite{SunRef11}, and ultra-sensitive detectors and filters (\textit {e.g.}, kinetic inductance detectors \cite{SunRef13}).

In high-field (MVm$^{-1}$-scale), high-operation-temperature (2\,--\,4.2\,K) SRF applications, surface oxide alterations and subsurface impurities significantly affect relevant cavity metrics of the industry-standard Nb (critical temperature $T_\mathrm{c}$\,=9.2\,K) \cite{SunRef21,SunRef40,SunRef7,SunRef8,SunRef23}. These foreign surface structures could also impact the next-generation SRF candidate niobium-tin (Nb$_3$Sn, $T_\mathrm{c}$\,=\,18\,K). The RF field only interacts with the near-surface layer of superconductors, \textit{i.e.}, $\sim$\,40\,nm in Nb and $\sim$\,100\,nm in Nb$_3$Sn, whereas the "dirty" native layer is typically several nanometers thick, with impurities appearing as deep as tens of nanometers (as observed in this work).
 
An ideal SRF surface could consist of either a dielectric capping layer or a normal-conducting proximity-coupled layer proposed by Gurevich and Kubo \cite{SunRef19,SunRef20}. The normal-conducting design could potentially utilize oxides of the desired thickness, structure, and electrical properties \cite{SunRef23}. One of the objectives of this work is to provide a map of the structures and depths of surface oxides for (i) \textit{in situ} heated Nb samples under ultra-high vacuum (UHV) and nitrogen (N$_2$) environments, (ii) 13 types of Nb coupon samples processed by acid/polishing, ozone, anodization, and heat treatments covering nearly all the protocols used in SRF preparations, (iii) Nb$_3$Sn coupons made by conventional vapor diffusion \cite{SunRef49} and the new electrochemical synthesis \cite{SunRef48}, and their \textit{in situ} vacuum heated samples.

Inspired by the theoretical surface design \cite{SunRef19,SunRef20}, Oseroff \cite{SunRef22,SunRef24} invented a thin-Au/Nb SRF surface, where a nanometer-thin normal-conducting gold (Au) layer could be relevant for losses and potentially beneficial if engineered with subnanometer thinness and specific electrical properties. The advantage of thinner surface oxides was also observed in tunneling experiments \cite{SunRef23}. This work is designed to support the construction of SRF surfaces by fundamentally understanding the nature of surface oxides and valence bonding with impurities and indicating electronic structures.  

Additionally, in low-field (Vm$^{-1}$-scale), low-operation-temperature (mK-scale) SRF applications (\textit {e.g.}, quantum qubits \cite{SunRef15}, cavity quantum electrodynamics (QED) \cite{SunRef15}, and SRF cavities used as QED architecture \cite{SunRef14}), oxide dielectrics create superconductor-insulator Josephson junction QED and enable the realization of logical operation states. However, amorphous oxides, which act as a two-level system, dominate dissipation and decoherence and hinder the preservation of logical states \cite{SunRef18,SunRef16,SunRef17}. This study in materials chemistry contributes to the investigation of loss mechanisms and the development of new interfaces for these emerging SRF devices.      

Numerous studies \cite{SunRef3,SunRef6,SunRef4,SunRef5,SunRef26,SunRef27,SunRef25,SunRef29,SunRef30,SunRef35,SunRef32,SunRef33,SunRef28} have investigated surface oxides and oxidization processes on Nb. The Nb surface is easily oxidized and presents signs of oxidization even with low oxygen exposures (0.2\,--\,2 langmuir) \cite{SunRef32,SunRef33}. Oxygen diffuses fairly fast in Nb with coefficients of 3\,$\times$\,10$^{-8}$\,cm$^2$/s at 800\,\textdegree C and 10$^{-21}$\,cm$^2$/s at 50\,\textdegree C \cite{SunRef34}. In recent years, native oxides on a high-purity Nb surface exposed to ambient pressure at room temperature (RT) are believed to comprise \textit{roughly} three layers in the following order: 1\,--\,6\,nm pentoxide (Nb$_2$O$_5$) / 0\,--\,2\,nm dioxide (NbO$_2$) / 0.3\,--\,1\,nm monoxide (NbO) / Nb (table~\ref{tbl:1}). In bulk studies, Nb$_2$O$_5$ exhibits amorphous, hexagonal, orthorhombic, tetragonal, and monoclinic structures at elevated temperatures \cite{SunRef36,SunRef37}, with 3.2\,--\,4.2\,eV band gaps being dielectric \cite{SunRef38}. NbO$_2$ has a distorted-rutile structure with a 0.7\,eV band gap being semiconducting (close to dielectric for the Nb SRF cavity operation). NbO has a vacancy-ordered rocksalt structure \cite{SunRef39} with a 1.4\,K $T_\mathrm{c}$ (normal-conducting at the cavity operation temperature). However, surface oxides do not fully follow their bulk properties. For example, surface Nb$_2$O$_5$ and NbO$_2$ decompose and disappear above 250\,--\,300\,\textdegree C, instead of transforming into another phase as the bulk behaves. Therefore, resolving accurate surface structures is critical to guiding the interpretation of loss mechanisms and the surface design of dielectric or proximity-coupled normal-conducting passivations.

Despite the large volume of literature (table~\ref{tbl:1})\cite{SunRef3,SunRef4,SunRef5,SunRef26,SunRef27,SunRef25,SunRef29,SunRef30,SunRef35,SunRef23}, significant variations exist across surface Nb oxides in terms of their native structures, depths, and decomposition products. Such discrepancies hinder a unified theory describing the SRF surface. For example, Eremeev \cite{SunRef21} observed inconsistent SRF responses between repeated treatments, despite \textit{in situ} modulation of surface oxide removal and re-growth. The probable causes include the unstable and non-uniform nature of surface oxides, the processing history, small variations between treatments (owing to the Nb surface's extremely high sensitivity/reactivity), and variations in instrument resolutions and analysis methods. These unclear results motivate revisiting the Nb surface oxide chemistry. 

\begin{table} [t!]
  \caption{Native surface oxides on Nb, thermal decomposition of oxides, bake-induced carbides, and subsurface impurities.}
  \label{tbl:1}
  \scalebox{0.6}{
  \begin{tabular}{p{7cm} p{7cm} p{3cm} p{3cm} p{3.2cm} p{2cm}}
    \hline
   Oxide structure and depth & Thermal decomposition of oxides & Bake-induced carbides & Subsurface impurities & Methods* & Ref. \\
    \hline
    2\,--\,6\,nm Nb$_2$O$_{5-x}$ / $\sim$\,1\,nm NbO$_{1-y}$ at (6\,--\,8)\,$\times$\,10$^{-10}$\,Torr & $<$\,1\,nm NbO$_{1-y}$ at 1850\,\textdegree C at 8\,$\times$\,10$^{-10}$\,Torr & -- & $\sim$\,50\,nm NbO$_{0.02}$ & Sputter-assisted XPS; ARXPS; AES & \cite{SunRef3,SunRef4,SunRef5,SunRef26} \\
    \hline
    Nb$_2$O$_{5}$ at 10$^{-10}$\,Torr & $<$\,1\,nm NbO at $>$\,325\,\textdegree C & -- & -- & XPS & \cite{SunRef27}\\
    \hline
    6\,nm Nb$_2$O$_{5}$ at 2\,$\times$\,10$^{-10}$\,Torr & 1.5\,nm Nb$_2$O$_5$ / 4.5\,nm NbO$_2$ at 280\,\textdegree C at 2\,$\times$\,10$^{-10}$\,Torr & C monolayers at $<$\,700\,\textdegree C; & -- & ARXPS & \cite{SunRef25}\\
    & $<$\,1\,nm NbO$_2$ / 1.5\,nm NbO at 380\,\textdegree C at 2\,$\times$\,10$^{-10}$\,Torr & NbC formation during cooling & &  & \\
    & $<$\,1\,nm NbO at $>$\,500\,\textdegree C at 2\,$\times$\,10$^{-10}$\,Torr & & &  & \\
     \hline
    3\,nm Nb$_2$O$_{5}$ / 1\,nm NbO$_x$ or 2\,nm Nb$_2$O$_{5}$ / NbO$_2$ / other oxides at 2\,$\times$\,10$^{-10}$\,Torr & $<$\,1\,nm Nb$_2$O at 250\,--275\,\textdegree C at 10$^{-9}$\,Torr & -- & -- & Synchrotron glancing-incidence XPS; ARXPS & \cite{SunRef29,SunRef30}\\
    \hline
    1\,nm Nb$_2$O$_{5}$ / 0.7\,nm NbO$_2$ / 0.3\,nm NbO at 8\,$\times$\,10$^{-10}$\,Torr & 0.8\,nm Nb$_2$O$_5$ / 0.7\,nm NbO$_2$ / 0.3\,nm NbO at 145\,\textdegree C at 8\,$\times$\,10$^{-10}$\,Torr & -- & $\sim$\,10\,nm O interstitials at 145\,\textdegree C at UHV & XRR; X-ray diffuse scattering; XPS & \cite{SunRef35}\\
    
    & 1\,nm NbO at 300\,\textdegree C at 8\,$\times$\,10$^{-10}$\,Torr & & & & \\ 
    \hline
    1\,--3\,nm Nb$_2$O$_{5}$ / 0.2\,--1\,nm NbO$_2$ / $\sim$\,1\,nm NbO at 8\,$\times$\,10$^{-9}$\,Torr & 2\,nm Nb$_2$O$_5$ / 1\,nm NbO$_2$ / $\sim$\,1\,nm NbO at 120\,\textdegree C at 8\,$\times$\,10$^{-9}$\,Torr & -- & -- & XRR; X-ray diffuse scattering;  & \cite{SunRef23}\\
    & $\sim$\,1\,nm NbO under 120\,\textdegree C 25\,mTorr N$_2$ "infusion" & & & XPS & \\
    & 0.5\,nm Nb$_2$O$_5$ / 0.2\,nm NbO$_2$ / $\sim$\,1\,nm NbO at 200\,\textdegree C at 8\,$\times$\,10$^{-9}$\,Torr & & & & \\
    & 0.2\,nm NbO$_2$ / $\sim$\,1\,nm NbO at 250\,\textdegree C at 8\,$\times$\,10$^{-9}$\,Torr & & & & \\
    & 0.2\,nm NbO$_2$ / $\sim$\,1\,nm NbO at 250\,\textdegree C 25\,mTorr N$_2$ "infusion" & & & & \\
    & 0.2\,nm NbO$_2$ / $\sim$\,1\,nm NbO / Nb$_x$N$_y$ under 500\,\textdegree C 25\,mTorr N$_2$ "infusion" & & & & \\
    & $\sim$\,1\,nm NbO at 800\,\textdegree C at 8\,$\times$\,10$^{-8}$\,Torr & & & & \\
    \hline
    1\,nm\,Nb$_2$O$_5$\,+\,Nb\,/\,2\,nm\,Nb$_2$O$_5$\,+\,NbO$_2$ +\,NbO\,+\,Nb\,/\,2\,nm\,Nb$_2$O$_5$\,+\,NbO$_2$\,+ NbO\,+\,NbO$_x$\,+\,Nb\,/\,1\,nm\,NbO$_2$\,+\,NbO +\,NbO$_x$\,+\,Nb\,/\,1\,nm\,NbO\,+\,NbO$_x$\,+\,Nb +\,NbC & 1\,nm\,Nb$_2$O$_5$\,+\,NbO$_2$\,+\,Nb\,/\,2\,nm\,Nb$_2$O$_5$ +\,NbO$_2$\,+\,NbO\,+\,Nb\,/\,2\,nm\,NbO$_2$\,+ NbO\,+\,NbO$_x$+\,Nb\,/\,2\,nm\,NbO\,+\,NbO$_x$ +\,Nb\,+\,NbC at 200\textdegree C at (0.7\,--\,2)\,$\times$\,10$^{-10}$\,Torr & NbC underneath the oxides \textit{in situ} at 200\textdegree C & $>$\,7\,nm O \& C at 200\textdegree C at UHV & Sputter-assisted XPS; ARXPS; STEM/EELS & EP'ed Nb (this work)\\   
    & No higher-order oxides \textit{in situ} at 500\,\textdegree C & 70\%\,--\,80\% NbC \textit{in situ} at 500\,\textdegree C &  $>$\,5\,nm O \& C at 500\textdegree C at UHV& & \\
    \hline
    1\,nm\,Nb$_2$O$_5$\,+\,NbO$_x$\,+\,Nb\,+\,NbC\,/\,1\,nm\,Nb$_2$O$_5$\,+\,NbO$_2$\,+\,NbO$_x$\,+\,Nb\,+\,NbC\,/ 6\,nm\,NbO\,+\,NbO$_x$\,+\,Nb\,+\,NbC\,\textit{in\,situ}\,at\,120\,\textdegree C\,at\,UHV\,(close\,to\,RT\,results) & & Up to 60\% NbC \textit{in situ} at 120\,\textdegree C, 300\,\textdegree C, 400\,\textdegree C, \& 800\,\textdegree C  & $>$\,40\,nm C \& $>$\,50\,nm O at 120\,\textdegree C, 300\,\textdegree C, 400\,\textdegree C, \& 800\,\textdegree C & Sputter-assisted XPS & 500\textdegree C pre-baked Nb (this work) \\
    \hline
    1\,nm\,Nb$_2$O$_5$\,+\,Nb\,+\,NbC & & Up to 50\% NbC & $>$\,30\,nm C \& O & Sputter-assisted XPS & 800\textdegree C pre-baked Nb (this work)\\
    \hline
    1\,nm\,Nb$_2$O$_5$\,+\,Nb\,+\,NbC\,/\,1\,nm\,Nb$_2$O$_5$\,+ NbO$_2$\,+\,NbO$_x$\,+\,Nb\,+\,NbC\,/\,1\,nm\,NbO$_2$ +\,NbO$_x$\,+\,Nb\,+\,NbC (embedded nitride nano-crystals) & & Up to 60\% NbC & $>$\,40\,nm C \& O & Sputter-assisted XPS; STEM /EELS & 800\textdegree C N$_2$-processed Nb (this work)\\
    \hline
    1\,nm\,Nb$_2$O$_5$\,+\,Nb\,+\,NbC & & Up to 20\% NbC & $>$\,20\,nm C \& O & Sputter-assisted XPS; STEM /EELS& N$_2$-processed + EP'ed (this work)\\
   \hline
\end{tabular}
}
*XPS: X-ray photoelectron spectroscopy; ARXPS: angle-resolved XPS; AES: Auger electron spectroscopy; XRR: X-ray reflectivity; STEM/EELS: scanning transmission electron microscopy with electron energy loss spectroscopy; UHV: ultra-high vacuum.
\end{table}

Important questions remain, further, of how multiple impurities (\textit{e.g.}, oxygen, O; carbon, C; and nitrogen, N) interact with Nb upon heating and affect subsequent oxidization. Impurity processing has recently become a key element of Nb SRF cavity treatments, producing record accelerating gradients and quality factors approaching the theoretical limit \cite{SunRef7,SunRef8}. These treatments involve N$_2$ processing (so-called "N$_2$ doping"\cite{SunRef8} and "N$_2$ infusion"\cite{SunRef7,SunRef42,SunRef43}) at 20\,--\,45\,mTorr N$_2$ pressures at low (120\,--\,160\,\textdegree C) \cite{SunRef7,SunRef42,SunRef43}, medium (300\,--\,400\,\textdegree C) \cite{SunRef44,SunRef41}, and high (800\,--\,1000\,\textdegree C)\cite{SunRef8} temperatures, or UHV baking similarly at low \cite{SunRef40} and medium \cite{SunRef45} temperatures. With claims of new recipes and new associated physics emerging, there is a need to provide an accurate picture of the processed Nb surface to build the foundation used for relevant theoretical modeling and process development. Secondary ion mass spectroscopy, SIMS, results have been widely reported \cite{SunRef7,SunRef43,SunRef46,SunRef41,SunRef48}. Confusion abounds concerning the extremely low nitrogen concentration ($\sim$\,0.01\,at.\%) in "N$_2$-infused" Nb, in contrast to 1\,--\,3 orders of magnitude higher oxygen and carbon concentrations (figure~S1) \cite{SunRef46}. Similarly, "N$_2$-doped" Nb, which requires post-electropolishing of the $\mu$m-thick surface, contains $<$\,(0.1\,--\,1)\,at.\% nitrogen and absolutely higher oxygen and carbon concentrations. Therefore, this work also aims to untangle the mystery of subsurface impurities in Nb in terms of their concentrations, structures, and valence distributions. 

Combining the implications of surface oxides and subsurface impurities, we extend the understanding of a desirable SRF surface that implies collective contributions. Previous studies have reported on the thinned oxide layer of N$_2$-processed Nb \cite{SunRef23,SunRef6} and the oxygen dissolution in UHV-baked Nb \cite{SunRef35}. Here, we expand on the efforts by considering multiple impurities and deconvoluting their multiplet structures, including carbides and suboxides. We provide a comprehensive understanding of structural identification and impurity bonding. To advance accuracy and throughput, our considerations span several aspects:

\begin{itemize}
      \item Sputter-assisted XPS and cross-sectional STEM/EELS are combined to calibrate depth-resolved chemical states; 
      \item Characteristics of XPS chemical states are identified in $>$500 \textit{in situ} samples using GENPLOT semi-automatic search algorithms to overcome the ambiguity of binding energy references;
      \item Empirical electronic structures (valence band electron distribution) are analyzed using XPS valence spectra fingerprints in the impurity region where hybridization is distinguishable \cite{SunRef47};
      \item Results are correlated between \textit{in situ} measurements during continuous heating at elevated temperatures and \textit{in situ} measurements of individual conditions (no heating history) versus coupon samples;
      \item Characterization is extended to Nb$_3$Sn, and the different oxide structures on vapor-diffused and electrochemically synthesized samples are highlighted. \cite{SunRef48}.
\end{itemize}

\section{Experimental Section}

\subsection{Nb and Nb$_3$Sn samples and treatments} 

High-purity ($>$\,300 residual-resistivity ratio), large-grain ($>$\,50\,mm), 3\,mm thick Nb plates were used. The Nb samples, unless otherwise specified, were polished following standard SRF protocols: first, buffered chemical polishing (BCP) using 1:1:1 hydrofluoric (HF, 48\%), nitric (70\%), and phosphoric (85\%) acids; second, 100\,$\mu$m electropolishing (EP) using 1:9 HF (48\%) and sulfuric (98\%, H$_2$SO$_4$) acids. The average surface roughness (R$_a$) of the EP'ed Nb was 40\,nm.

As summarized in table~\ref{tbl:2}, treatments of interest include (i) acid-related treatments such as BCP, EP, and 2\% HF soak; (ii) heat-related treatments such as 800\,\textdegree C UHV (10$^{-10}$ torr) baking, "N$_2$ doping", and "N$_2$ infusion"; and (iii) environment-related treatments such as modulated 20\%:80\% O$_2$\,/\,N$_2$ soak, ozone processing (Jelight 144AX), and exposure under ambient conditions for varying periods. The typical "2/800 N$_2$ doping" recipe involved exposing the Nb to 45\,mTorr N$_2$ at 800\,\textdegree C for 2\,min, followed by a 5\,$\mu$m EP removal \cite{SunRef8,SunRef43}. Except for ozone exposure and anodization, these treatments adhered to the typical procedures employed in Nb cavity preparation \cite{SunRef46}. The reactive ozone treatment was designed to modify surface oxides, while anodization is a typical treatment of Nb substrates utilized prior to the Sn-vapor diffusion process to produce Nb$_3$Sn.

Approximately 2\,$\mu$m thick Nb$_3$Sn films were produced using two methods: (i) Sn-vapor diffusion \cite{SunRef49} and (ii) a recently developed process that involves combining electrochemical Sn deposits with thermal conversion to a smoother Nb$_3$Sn \cite{SunRef48}. The average surface roughnesses (R$_a$) for these two methods were 300\,nm and 40\,nm, respectively. 

After undergoing treatments, the samples were stored in a nitrogen-flowed glovebox and transported in cleanroom plastic bags before any characterization. 

\begin{table} [tb]
  \caption{Summary of acid/polishing, ozone, anodization, and heat treatments on Nb coupons.}
  \label{tbl:2}
 \scalebox{0.8}{
  \begin{tabular}{p{2cm}p{2cm}p{2cm}p{2cm}p{2cm}p{2cm}p{2cm}p{2cm}}
    \hline
   Sample \# & SMP1 & SMP1 & SMP1 & SMP1 & SMP2 & SMP3 & SMP3 \\
   Step \# & STP1 & STP2 & STP3 & STP4 & -- & STP1 & STP2  \\
   Treatment & BCP & HF soak 30\,min & 2nd BCP 30\,min & Ozone exposure 3\,d & 100\,$\mu$m EP & 100\,$\mu$m EP & RT O$_2$/N$_2$ flow $>$\,1\,d \\
   Pre-treatment & N/A & N/A & N/A & N/A & BCP & No prior BCP & N/A  \\
   Post air exposure & $>$\,6\,mo & $<$\,3\,d & $<$\,3\,d & $<$\,3\,d & $<$\,3\,d & $<$\,3\,d & $<$\,3\,d \\
    \hline
   Sample \# & SMP4 & SMP5 & SMP6 & SMP6 & SMP7 & SMP8 & SMP9 \\
   Step \#  & -- & -- & STP1 & STP2 & -- & -- & -- \\
   Treatment & 800\,\textdegree C 5\,h baking + RT O$_2$/N$_2$ flow $>$\,1\,d & “2/800” N$_2$ doping + 5 $\mu$m light EP & 100\,$\mu$m EP + RT N$_2$ storage & “2/800” N$_2$ doping & 100\,$\mu$m EP + RT N$_2$ storage & 160\,\textdegree C N$_2$ infusion 96\,h & Anodization \\
   Pre-treatment & BCP + EP & BCP + EP + 800\,\textdegree C 5\,h baking & No prior BCP & 800\,\textdegree C 5\,h baking & N/A & BCP + EP + 800\,\textdegree C 5\,h baking & EP \\
   Post air exposure & $>$\,6\,mo & $>$\,6\,mo & $<$\,3\,d & $<$\,3\,d & $<$\,3\,d & $>$\,6\,mo & $<$\,3\,d \\
     \hline
\end{tabular}
}
\end{table}

\subsection{XPS and STEM/EELS characterizations} 

Quantitative determination of chemical states, elemental compositions, and valence distributions was primarily achieved through XPS depth profiling, supported by cross-sectional STEM/EELS.

Three XPS systems were employed for this study, namely a PHI Versaprobe III, a Surface Science Instruments SSX-100 ESCA Spectrometer, and a Thermo Fisher Nexsa G2, with binding energy calibrations cross-referenced using EP'ed Nb. Comparing the monoatomic argon and cluster ion sputtering results from Nexsa (figure~S2(a)), we found mono-sputtering provided effective surface removal and minimal oxide residuals. The non-destructive ARXPS results (figure~S2(b)) were consistent with the structural evolution of the oxide layers as resolved by sputter-assisted XPS. Although preferential sputtering, atomic mixing, and the effect of roughness can occur \cite{SunRef50}, sputter-assisted XPS allows probing the impurity region, sitting as deep as tens of nanometers (beyond the $<$\,5\,nm capability of ARXPS). Also, understanding chemical bonding (XPS peak position) is the priority over other tasks in this work, so mono-sputter-assisted depth profiling was primarily used.

\subsubsection{\textit{In situ} XPS with heating stage and reaction cells} 

To capture critical changes during UHV and N$_2$ bakings, as well as after subsequent air exposure, \textit{in situ} investigations were performed in three sets using the PHI Versaprobe XPS instrument with a heating stage. First, EP'ed Nb was repeatedly heated at 200\,\textdegree C, 500\,\textdegree C, 120\,\textdegree C, 300\,\textdegree C, 400\,\textdegree C, and 800\,\textdegree C at (0.7\,--\,2)\,$\times$\,10$^{-10}$\,Torr UHV for 30\,min, with intentional ambient air exposure for 10\,--\,24\,h between measurements to imitate SRF cavity treatments. Depth profiles were collected \textit{in situ} during heating at the same temperature as the 30\,min pre-heating, except for 800\,\textdegree C processing. For this, an adjacent UHV reaction chamber (8\,$\times$\,10$^{-9}$\,Torr) was required. The sample processed at 800\,\textdegree C was transferred directly into the analysis chamber, without exposure to the atmosphere, under 8\,$\times$\,10$^{-10}$\,Torr and measured at 500\,\textdegree C, the highest temperature available in the XPS analysis chamber. Additionally, the 500\,\textdegree C data were collected several times to confirm the observation of high carbon concentrations induced, which increased the number of sample heating cycles. Second, fresh-EP'ed Nb was processed to resemble the 800\,\textdegree C "N$_2$ doping" procedures, which involved subjecting the Nb to 800\,\textdegree C N$_2$ processing at 1\,mTorr for 90\,min (the same amount of N$_2$ exposure as "N$_2$ doping"), followed by re-exposure to air and a final 5\,$\mu$m EP, with depth profiling conducted between each procedure. Third, a reference 800\,\textdegree C UHV baked sample was examined immediately after baking and after air exposure, which provided comparisons with "N$_2$ doping" and avoided effects induced by the heating history. The samples were tightly mounted on the heating stage, and the stage temperature was calibrated and showed negligible variation (typically $<$ $+/-$ 1 degree).

High-resolution spectra of Nb 3d, Nb 3p, O 1s, C 1s, and valence photoelectrons were probed using a 100 $\mu$m monochromatic Al k-alpha X-ray (1486.6\,eV) beam. After optimization (figure~S3), the scan parameters were set to 45\textdegree~emission angle, 26\,eV pass energy, 50\,ms/step, and up to 60 sweeps with the dual-neutralization on, resulting in a 0.3\,--0.6\,eV FWHM resolution for Nb subpeaks. For depth profiles, a 3\,keV Ar$^+$ beam was rastered over 2\,$\times$\,2\,mm$^2$ area with Zalar rotation. The sputtering rate of 1.6\,\AA/s was determined using a SiO$_2$ standard and compared with the cross-sectional STEM result. Depth profiles were collected with 6\,s intervals at the surface region and 12\,--\,60\,s intervals toward the bulk.

Vapor-diffused Nb$_3$Sn was \textit{in situ} measured at RT, 200\,\textdegree C, and 500\,\textdegree C under continuous heating, without exposure to air, using the same parameters as for Nb measurements.  

\subsubsection{Coupon inspections} 

To obtain depth profiles of oxides on the coupons, we conducted survey scans of representative Nb samples listed in table~\ref{tbl:2}, as well as two types of Nb$_3$Sn samples, using the SSX-100 XPS instrument. Monochromatic Al k-alpha X-ray (1486.6\,eV) photoelectrons were collected under a 10$^{-9}$\,Torr vacuum from an 800\,$\mu$m analysis spot with a 55\textdegree~emission angle. The scan parameters were set to 150\,eV pass energy, 1\,eV step size, and 100\,s/step. For the depth profile, a 4\,kV Ar$^+$ beam with a spot size of $\sim$\,5\,mm was rastered over a 2\,$\times$\,4 mm$^2$ area.

We used an FEI/Thermo Fisher Titan Themis STEM with beam energy up to 300\,kV and resolution down to 0.08\,nm to image the surface cross-sections of EP'ed and "N$_2$ doped" Nb and vapor-diffused Nb$_3$Sn. The equipped EELS was utilized to analyze the chemical states as a function of depth at local regions. The cross-sectional specimens were prepared using a Thermo Fisher Helios G4 UX focused ion beam.

\section{Results and Discussion}

\subsection{Native oxides on Nb at RT}

\begin{figure*}[b!]
\centering
\includegraphics[width= \linewidth]{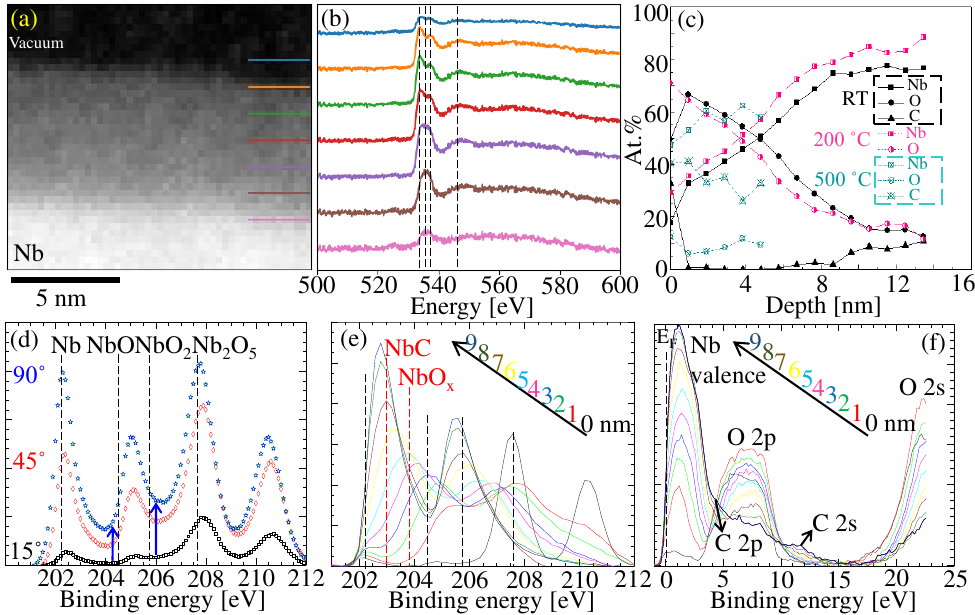}
\caption{Native oxides on EP'ed Nb at RT: (a) Cross-sectional STEM image showing surface oxides on Nb. (b) EELS O-K edge spectra at various depths as labeled in (a). (c) Concentration depth profiles of Nb, O, and C collected by XPS at RT (and during \textit{in situ} heating at 200\textdegree C and 500\textdegree C). (d) ARXPS spectra collected at take-off angles of 15\textdegree, 45\textdegree, and 90\textdegree. (e) Nb 3d and (f) valence XPS spectra as a function of depth. The intensity units in (b) and (d\,--\,f) are arbitrary.}
\label{SunFig1}
\end{figure*}

The cross-sectional STEM image (figure~\ref{SunFig1}(a)) reveals an amorphous oxide region of $\sim$\,7\,nm on the surface of EP'ed Nb (measured at RT). Due to the limitations of each technique, structural deconvolution requires the combination and correlation of findings from EELS, ARXPS, and sputter-assisted XPS. The structure consists of Nb$_2$O$_5$ mixed with metallic Nb at the outermost layer, along with hydroxide and organics adsorbed. The next layer, which is 4\,--\,5\,nm thick, comprises a mixture of Nb$_2$O$_5$, NbO$_2$, NbO, NbO$_x$ (suboxide), and Nb. The following layer, which is 1\,--\,2\,nm thick, is a mixture of NbO, NbO$_x$, and Nb, with NbO$_x$ continuing and NbC appearing deeper. The evidence of these identifications is detailed below.  

EELS spectra (figure~\ref{SunFig1}(b)), taken at the locations indicated in figure~\ref{SunFig1}(a), reveal two distinct regions with thicknesses of $\sim$\,5\,nm and $\sim$\,2\,nm, respectively. Figure~S4 summarizes the EELS fingerprints collected from Nb$_2$O$_5$, NbO$_2$, and NbO in the literature \cite{SunRef31}, showing that the Nb$_2$O$_5$ and NbO$_2$ features have close energy states while NbO exhibits distinguishable energies near 535 eV and 548 eV. Based on this, the EELS data suggests that the first 5\,nm of the oxide region is mixed with Nb$_2$O$_5$/NbO$_2$ and NbO, while the second 2\,nm is primarily NbO. This local probe substantiates the mixing composition within each layer of surface oxides.   

The XPS concentration profile (figure~\ref{SunFig1}(c)) also shows an oxide region of $\sim$\,7\,nm in thickness, followed by deeper oxygen and carbon impurities. Additionally, contamination was observed at the outermost surface of all coupon samples studied, likely due to methanol cleaning. The deconvoluted structure of the hydroxyl, methoxy, carbonyl, and carboxyl dangling bonds is depicted in figure~S5. 

ARXPS measurements, taken at a 15\textdegree\,take-off angle (figure~\ref{SunFig1}(d)), strongly indicate the presence of metallic Nb mixed within Nb$_2$O$_5$, as previously observed in a 2\textdegree\, grazing angle measurement \cite{SunRef30}. Deeper probing at higher angles between 45\textdegree\,and 90\textdegree\,detected the NbO$_2$ and NbO signals. This near-grazing measurement supports the observation of metallic Nb within the oxide layers in sputter-assisted XPS.  

\begin{figure*}[h]
\centering
\includegraphics[width= \linewidth]{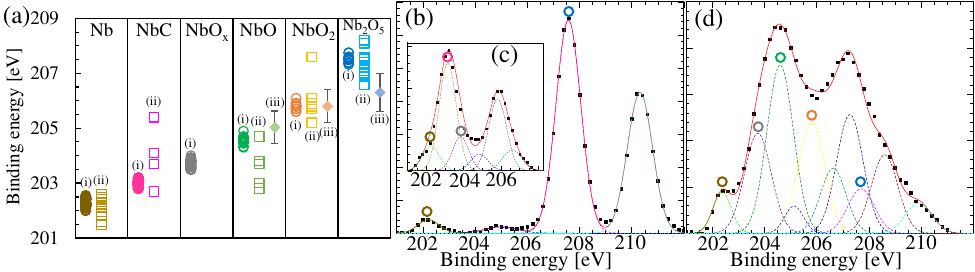}
\caption{Nb 3d XPS binding energy analysis: (a) Binding energies of Nb in different chemical states (metallic Nb, carbide, suboxide, monoxide, dioxide, and pentoxide) determined using (i) this work, (ii) the NIST reference database \cite{SunRef51} and (iii) EELS chemical shifts \cite{SunRef31}. (b,c,d) Representative XPS spectra of EP'ed Nb showing characteristic peaks used for peak fitting: (b) at RT without sputtering, (c) at RT after sputtering 3\,nm, and (d) at 500\,\textdegree C without sputtering. In (b\,--\,d), the labeled circles are color-coded to correspond with the identifications in (a), and the intensity units are arbitrary.}
\label{SunFig2}
\end{figure*}
 
The sputter-assisted XPS spectra in figure~\ref{SunFig1}(e) clearly show the evolution of structural changes as a function of depth. To deconvolute this complex system with diverging references, we utilized GENPLOT semi-auto searching in vast amounts of data to elucidate the binding energies of all components present on the Nb surface, as summarized in figure~\ref{SunFig2}. The identification is further corroborated by concurrent features observed in the valence spectra, such as C 2s (figure~\ref{SunFig1}(f)). We have identified Nb at 202.2\,$\pm$\,0.1\,eV, NbC at 203\,$\pm$\,0.1\,eV, NbO$_x$ at 203.8\,$\pm$\,0.1\,eV, NbO at 204.5\,$\pm$\,0.1\,eV, NbO$_2$ at 205.8\,$\pm$\,0.1\,eV, and Nb$_2$O$_5$ at 207.5\,$\pm$\,0.1\,eV. All subpeaks, except for NbO$_x$, exhibit their characteristic peaks, as exemplified in figures~\ref{SunFig2}(b)--(d); the incorporation of NbO$_x$ is nevertheless necessary to fit the spectra in 375 samples. With the support of EELS and ARXPS analyses, we can trust the validity of these observations, although sputtering effects may slightly influence the results.  

We emphasize that metallic Nb coexists with amorphous oxides exhibiting a mixing feature, likely involving non-stoichiometry and defective energy states near the Fermi level. Both findings are consistent with the observation of electron populations at the Fermi level in the valence spectra of surface layers, while bulk signals contribute to the Fermi electrons at deeper layers, as shown in figure~\ref{SunFig1}(f). Our observations strongly suggest that the entire $\sim$\,7\,nm surface oxide region is conductive, most likely normal-conducting, and plays a crucial role in determining the surface properties of superconducting Nb. Further identification of sub-nanometer local amorphous structures is helpful to confirm this finding.  

\subsection{Effects of UHV baking temperature on Nb}

\begin{figure*}[h]
\centering
\includegraphics[width= \linewidth]{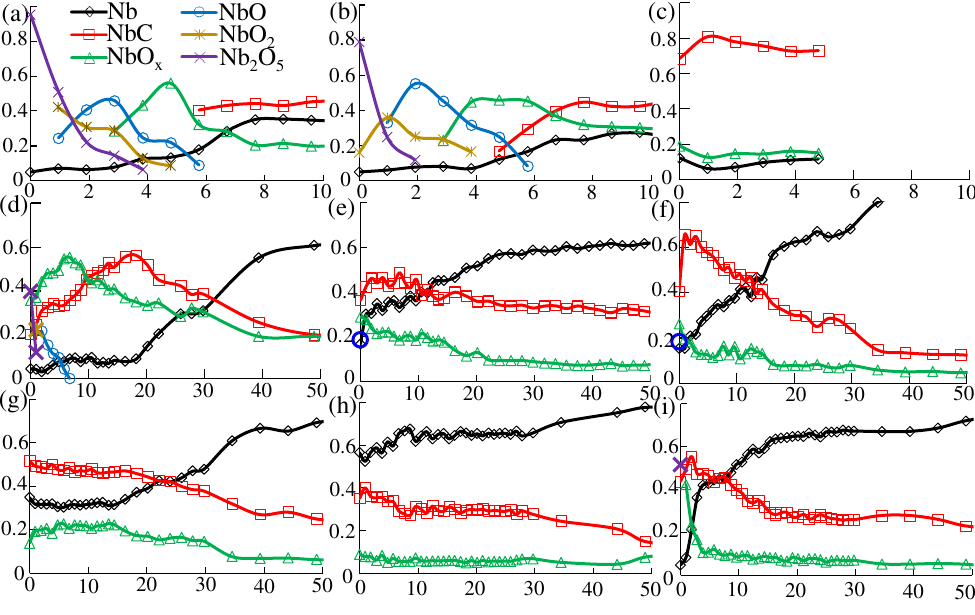}
\caption{Structural profiles of UHV-baked Nb: Comparative proportion of different Nb motifs resolved by XPS peak fitting of Nb 3d spectra, plotted as a function of depth in nm. The spectra were taken \textit{in situ} at (a) RT, (b) 200\textdegree C, (c) 500\textdegree C, (d) 120\textdegree C, (e) 300\textdegree C, (f) 400\textdegree C, and (g) 800\textdegree C, in the indicated sequence, with air exposure between measurements. A reference sample was (h) baked directly at 800\textdegree C and then (i) exposed to air. The expected fitting residue is between 5\%\,--\,10\,\%.}
\label{SunFig5}
\end{figure*}

\textit{In situ} XPS measurements were taken on Nb at various temperatures, with deliberate air exposure between each measurement. A high concentration of carbon appeared in the sample at 500\textdegree C (figure~\ref{SunFig1}(c)), which crucially affected the subsequent oxidization and impurity bonding. Thus, we divide the results into two groups: (i) the effects of UHV baking temperature on EP'ed Nb, with a heating history of RT, 200\textdegree C, and 500\textdegree C, and (ii) the effects of UHV baking temperature on high-temperature baked Nb (close to the 800\textdegree C "outgassing" baking condition), with a history of 500\textdegree C, 120\textdegree C, 300\textdegree C, 400\textdegree C, and 800\textdegree C. The complete Nb 3d and valence spectra are included in figures~S6 and S7, respectively.  

\subsubsection{Effects of UHV baking temperature on EP'ed Nb}

Figure~\ref{SunFig5} presents a map of the surface oxide and carbide motifs for all baking conditions. When heated at 200\textdegree C, the oxides undergo slight reconstruction and thickness reduction, while the layout remains similar to RT, which is consistent with the decomposition threshold at $\sim$\,250\,--\,300\,\textdegree C reported in the literature \cite{SunRef25,SunRef6,SunRef35}. However, upon heating at 500\textdegree C, nearly 80\% of the surface structures convert to carbides, with the higher-order oxides disappearing. The binding energy of the main peak in the spectra shifts to 203\,eV (figure~\ref{SunFig2}(c)). These carbides significantly alter the subsequent oxidization process and the diffusion of impurities into the bulk upon re-exposure to air.   

\subsubsection{Effects of UHV baking temperature on high-temperature baked Nb}

\begin{figure*}[h]
\centering
\includegraphics[width= \linewidth]{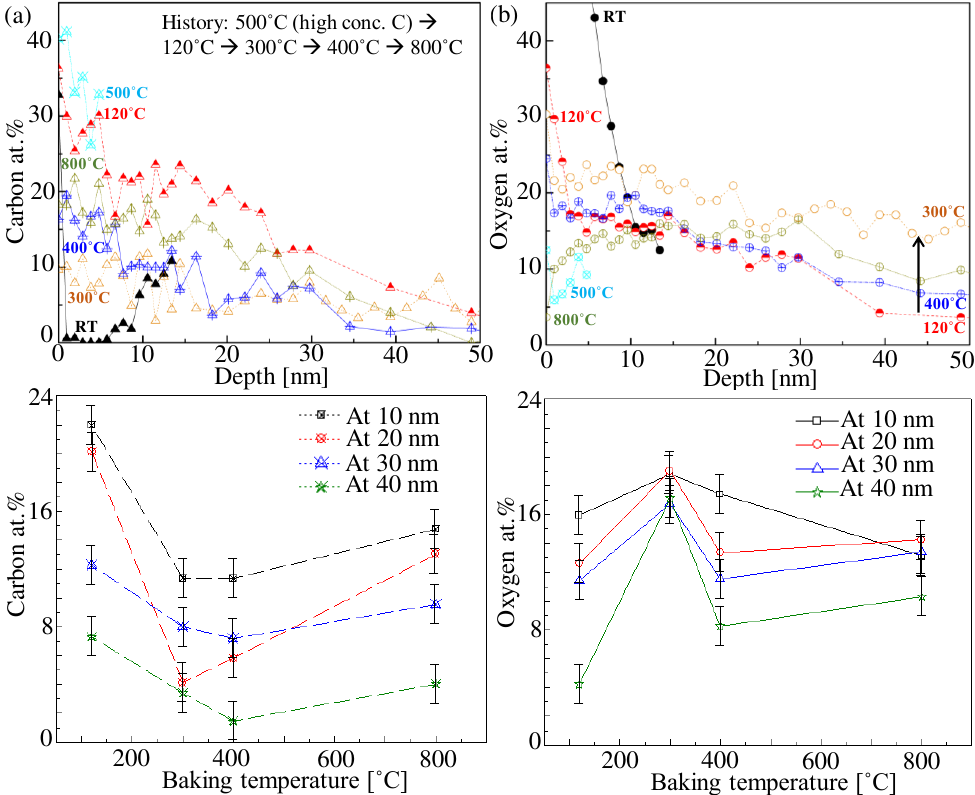}
\caption{Composition profiles of UHV-baked Nb: (a) Carbon and (b) oxygen concentrations as a function of depth measured \textit{in situ} at different temperatures, following a heating history of 500\textdegree C, 120\textdegree C, 300\textdegree C, 400\textdegree C, and 800\textdegree C. (c) Carbon and (d) oxygen concentrations at selective depths as a function of baking temperature. The starting RT reference for this experiment was an EP'ed sample.}
\label{SunFig3}
\end{figure*}

Figures~\ref{SunFig3}(a) and (b) illustrate the changes in carbon and oxygen concentrations with baking temperature after the 500\textdegree C pre-baking. A thick impurity region is observed with a carbon thickness of 20\,--\,40\,nm and an oxygen thickness exceeding 40\,nm (using a conservative residue signal threshold of 5\,at.\%). We argue that the influence from chamber gas residues is negligible since nitrogen consistently remains below the detection limit ($<$\,1000\,ppm), as measured by residual gas analysis. We also argue that the sputtering effect is minor since the RT data is consistent with the results of STEM imaging. However, we find that the heating history affects the impurity profile. For instance, directly heating the EP'ed sample to 800\textdegree C results in a carbon thickness of 20\,nm and an oxygen thickness of $\sim$\,30\,nm. 

The comparison between the 200\textdegree C (figures~\ref{SunFig5}(b) and S6(b)) and 120\textdegree C (figures~\ref{SunFig5}(d) and S6(d)) data, both temperatures of which are below the decomposition threshold, implies that the initial surface plays a critical role in oxidation. The 200\textdegree C oxides form on an EP'ed surface, whereas the 120\textdegree C oxides grow on a surface populated with carbides from a 500\textdegree C pre-baking. The presence of carbides essentially minimizes the generation of higher-order oxides. The transition metal (\textit{e.g.}, Nb) favors dissociation into chemisorbed oxygen, and any foreign phases interrupt the self-passivating oxides, which allows for oxygen (and carbon) dissolution into the subsurface \cite{SunRef52}. Considering the low equilibrium O and C solubilities in Nb ($<<$\,1\,at.\% even at 800\textdegree C \cite{SunRef53,SunRef54}), a pertinent question is regarding the oversaturation of oxygen and carbon in the subsurface.

Although we adopt the buried oxide (or precipitate) hypothesis and treat them as carbides and suboxides in this work, there are arguments regarding the accurate picture \cite{SunRef52}. Our other work observed the existence of carbides in the subsurface \cite{SunRef55}, while our valence analyses (below) suggest that these subsurface impurities are not likely to form a specific phase, especially for the suboxide. Nevertheless, our data indicate that the UHV baking mechanism in the SRF field mainly takes advantage of preceding carbide formation during the high-temperature treatment, \textit{e.g.}, 800\,\textdegree C outgassing. This process surpasses the limiting rate during impurity dissolution, resulting in a mixture of carbides and suboxides, in addition to varying oxide thickness. 

Upon heating to 300\textdegree C, oxygen accumulates on the surface over carbon in concentration (figures~\ref{SunFig3}(c) and (d)), which matches the onset of native oxide decomposition. However, carbides continue to dominate over suboxides (figure~\ref{SunFig5}(e)). Baking at 400\textdegree C yields similar composition and motif proportion profiles compared to baking at 300\textdegree C, while carbide formation is intensified at the relatively surface region with a 60\% proportion compared to the bulk region with a 20\% proportion, as shown in figure~\ref{SunFig5}(f). At 800\textdegree C (figure~\ref{SunFig3}(d)), there is a surface depletion of oxygen, with the proportion of pure Nb increasing to above 30\% at the 20\,nm surface in addition to 50\% carbides (figure~\ref{SunFig5}(g)). For reference, the sample that was directly heated to 800\textdegree C (figure~\ref{SunFig5}(h)) has 30\%\,--\,40\% carbides and 60\%\,--70\,\% pure Nb on the surface with minimal suboxides. However, air exposure of this sample (figure~\ref{SunFig5}(i)) immediately introduces higher numbers of carbides and suboxides within 20\,nm, along with a one-nanometer-thin higher-order oxide region, demonstrating the higher-order oxide formation limitation and impurity uptake due to carbide formation. 

\begin{figure*}[b!]
\centering
\includegraphics[width= \linewidth]{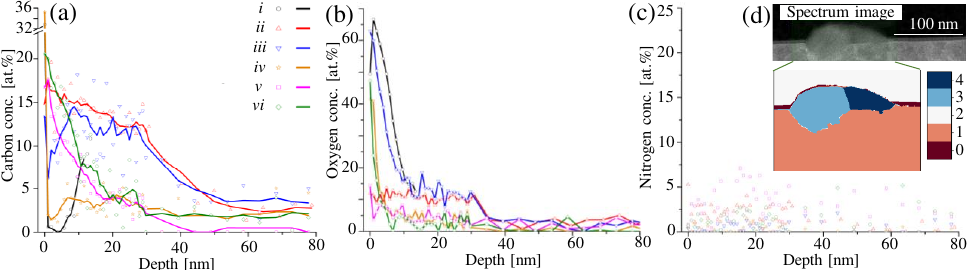}
\caption{Composition profiles of N$_2$-processed Nb: Depth profiles of (a) carbon, (b) oxygen, and (c) nitrogen concentrations for different processing conditions: (i) after pre-EP, (ii) during \textit{in situ} N$_2$ processing at 800\textdegree C ("N$_2$ doping"), (iii) after air exposure, and (iv) after re-EP (light). A reference sample was subjected to (v) UHV baking at 800\textdegree C and then (vi) air exposure. (d) Cross-sectional phase-contrast STEM image of the N$_2$-processed coupon (condition iii), showing surface nitride nano-grains (labeled "3" and "4") and an oxide layer (labeled "0").}
\label{SunFig8}
\end{figure*}

Additionally, upon inspecting the O 1s spectra (\textit{e.g.}, figure~S8), we identify two types of oxygen-related motifs, assigned to O$^{2-}$ and O$^{-}$. As illustrated in figure~S9, the change in the relative ratio of O$^{-}$ likely correlates with the carbon concentration as a function of depth, indicating a possible interaction between carbon and certain oxygen species.

\subsection{Effects of N$_2$ processing on Nb}
\begin{figure*}[b!]
\centering
\includegraphics[width= \linewidth]{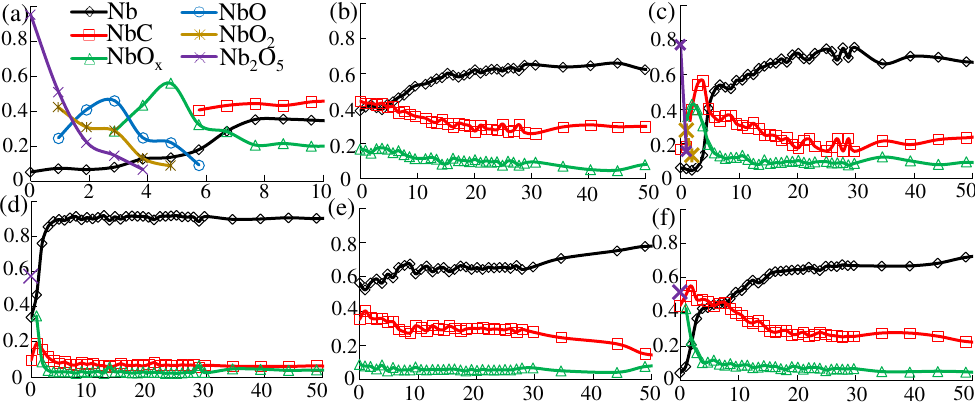}
\caption{Structural profiles of N$_2$-processed Nb: Comparative proportion of different Nb motifs as a function of depth in nm. The spectra were taken (a) after pre-EP, (b) during \textit{in situ} N$_2$ processing at 800\textdegree C ("N$_2$ doping"), (c) after air exposure, and (d) after re-EP (light) treatments. A reference sample was subjected to (e) UHV baking at 800\textdegree C and then (f) air exposure. The expected fitting residue is between 5\%\,--\,10\,\%.}
\label{SunFig6}
\end{figure*}
Two sets of EP'ed Nb were processed: one at 800\textdegree C with 5.4\,$\times$\,10$^6$ langmuir N$_2$ exposure to simulate "N$_2$ doping", and the other under UHV as a reference. The complete Nb 3d and valence spectra are included in figures~S10 and S11, respectively. 

\begin{figure*}[t!]
\centering
\includegraphics[width= \linewidth]{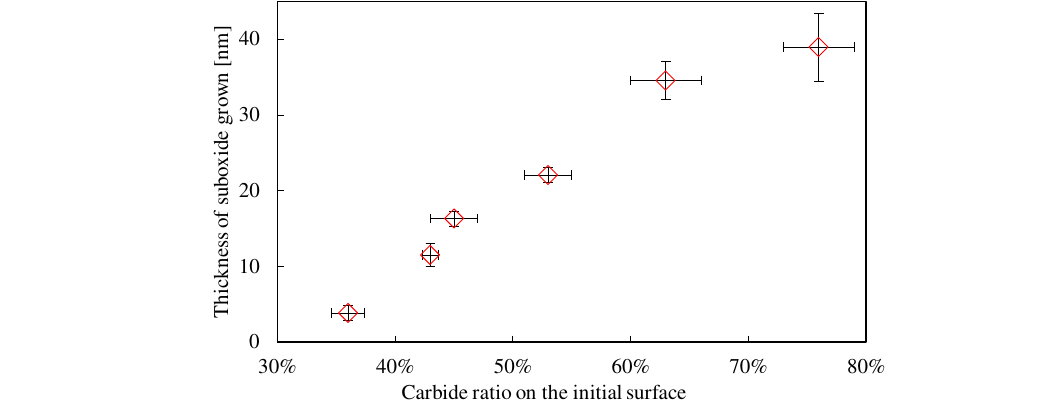}
\caption{Effect of second-phase formation: The correlation between the thickness of suboxide layers and the varying ratio of carbides on the surface where suboxides are formed. The data was obtained through \textit{in situ} UHV baking (history: 500\textdegree C, 120\textdegree C, 300\textdegree C, 400\textdegree C, and 800\textdegree C), \textit{in situ} direct 800\textdegree C UHV baking and N$_2$ processing, and their subsequent air exposure.} 
\label{SunFig14}
\end{figure*}

Figure~\ref{SunFig8} shows that both \textit{in situ} conditions introduce significant amounts of carbon and oxygen, but nitrogen is below the detection limit in all samples studied ($<$\,1000\,ppm), even for the "N$_2$ doping" condition, despite the observation of two nitride nano-grains during STEM inspection. Structural deconvolution reveals that the N$_2$ processed sample (figure~\ref{SunFig6}(b)) contains slightly higher levels of carbides and suboxides than the UHV baked sample (figure~\ref{SunFig6}(e)). Upon air exposure (figure~\ref{SunFig6}(c)), the suboxide layer on the N$_2$-processed sample surface substantially increases, along with more higher-order oxides. In contrast, the UHV baking reference (figure~\ref{SunFig6}(f)) only exhibits a 3\,nm suboxide layer with proportions exceeding 10\% and 1\,nm with higher-order oxides. These findings, together with the UHV baking data, prove that the thickness of the suboxide layer, grown on surfaces populated with carbides, is proportional to the ratio of carbides, as illustrated in figure~\ref{SunFig14}. Specifically, an initial surface with a higher number of carbides results in a larger amount of suboxide growth. The suboxide layer profile, including its peak concentration and location, determines the type and thickness of higher-order oxides formed on top of the suboxide layer. 


To achieve high-performance SRF cavities, a 5\,$\mu$m (light) EP treatment is commonly employed after N$_2$ processing to remove any nitrides. We have examined the surface profile following air exposure of the EP'ed, N$_2$-processed surface. Our study shows that this surface (figure~\ref{SunFig6}(d)) contains the lowest quantities of carbides and suboxides, alongside the thinnest higher-order oxides. Notably, variations in the oxide profile exist between EP'ed, N$_2$-processed Nb and EP'ed Nb. While the electrochemical mechanism is beyond the scope of this study, our focus centers on valence analyses to demonstrate the modification of electron populations in orbitals resulting from these treatments. 

\subsection{Valence analyses}

\begin{figure*}[h]
\centering
\includegraphics[width= \linewidth]{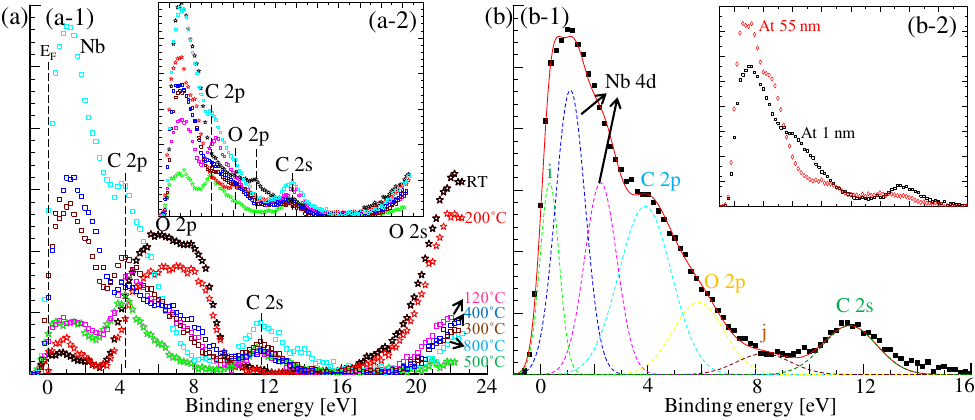}
\caption{Valence spectral analysis in UHV-baked Nb: (a) Comparison of valence spectra taken  \textit{in situ} at different temperatures, following the sample's heating history of RT, 200\textdegree C, 500\textdegree C, 120\textdegree C, 300\textdegree C, 400\textdegree C, and 800\textdegree C, at depths of (a-1) 0\,nm and (a-2) 9\,nm (or the maximum depth probed). (b-1) Example of valence peak fitting for spectra taken at 1\,nm during \textit{in situ} baking at 800\textdegree C, and (b-2) comparison with spectra taken at 55\,nm, demonstrating characteristic features. The binding energy scales in (a-2) and (b-2) are equivalent to those in (a-1) and (b-1), respectively.}
\label{SunFig4}
\end{figure*}

The valence bonding in carbides and oxides typically involves hybridization, which causes overlapping energy levels in momentum space. This makes it challenging to distinguish the contribution of electronic density of states in each orbital to the overall electronic structure. First-principles calculations \cite{SunRef56,SunRef57} have reported that higher-order Nb oxides exhibit strong hybridization (as observed in figure~\ref{SunFig1}(f)). Conversely, lower-order Nb oxides and carbides show weaker hybridization and more distinguishable orbital occupation \cite{SunRef56,SunRef58}. 

When the transition metal d-orbitals are partially filled with low numbers of electrons ($<$\,3), these orbitals and the O or C orbitals are roughly separated\cite{SunRef64,SunRef65}. XPS studies (figure~S12) have revealed that pure Nb displays three characteristic peaks at 0\,--\,3\,eV, with the majority contribution coming from the Nb 4d orbital \cite{SunRef59,SunRef60,SunRef61}. Exposure to oxygen results in a peak near 6.5\,eV, mainly from O 2p \cite{SunRef32}. NbC shows two peaks: one near 4.4\,eV, mainly from C 2p, and another near 11\,eV from C 2s \cite{SunRef63,SunRef60,SunRef58}. 

\begin{figure*}[h]
\centering
\includegraphics[width= \linewidth]{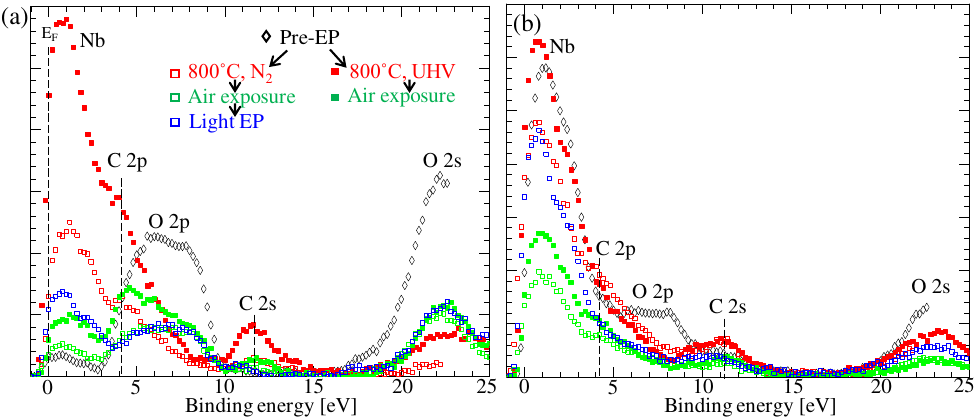}
\caption{Valence spectral analysis in N$_2$-processed Nb: Comparison of valence spectra taken after pre-EP, during \textit{in situ} N$_2$ processing at 800\textdegree C ("N$_2$ doping"), after air exposure, and after re-EP (light): (a) at 0\,nm and (b) at 9\,nm. A reference sample was baked at 800\textdegree C under UHV and then exposed to air. These results are re-plotted as figures~\ref{SunFig6}(e) and (f) to facilitate comparison. The intensity units in (a) and (b) are arbitrary.}
\label{SunFig9}
\end{figure*}

Figures~\ref{SunFig4} and \ref{SunFig9} demonstrate that we repeatedly and clearly identify these characteristic peaks in over 500 valence spectra, except for those with higher-order oxides. To deconvolute the spectra, we perform peak fitting using 7 Gaussian peaks, as exemplified in figure~\ref{SunFig4}(b). The sub-peak denoted "i" near the Fermi edge may include surface states and a large contribution from Nb 5s (and possibly 5p if hybridized). Another small sub-peak denoted "j" is likely a hybridization product of Nb-O, as indicated by higher-order oxides. Consequently, these sub-peaks are not included in the quantitative calculations.

\begin{figure*}[h]
\centering
\includegraphics[width= \linewidth]{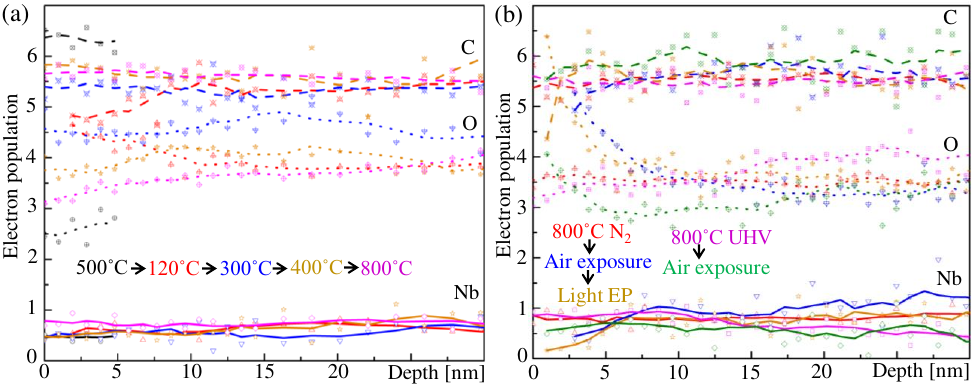}
\caption{Electron population analysis in UHV-baked and N$_2$-processed Nb: Electron population as a function of depth in nm for Nb (solid line), O (dotted), and C (dashed) orbitals. (a) The \textit{in situ} UHV baked sample, following a heating history of 500\textdegree C (black square), 120\textdegree C (red up-triangle), 300\textdegree C (blue down-triangle), 400\textdegree C (orange star), and 800\textdegree C (purple diamond). (b) The \textit{in situ} 800\textdegree C N$_2$ processed ("N$_2$ doped") sample (red up-triangle), followed by air exposure (blue down-triangle) and light EP (orange star). A reference sample was directly baked at 800\textdegree C under UHV (purple diamond) and then exposed to air (green square).}
\label{SunFig7}
\end{figure*}

We calculate the electron population at each bonding orbital through realistic normalization, such as in Nb ($N_{\mathrm{Nb}}$), by \cite{SunRef65,SunRef66,SunRef67}

\begin{eqnarray}
 N_{\mathrm{Nb}}~=~{\frac{A_{\mathrm{Nb}}~/~[{\delta_{\mathrm{Nb}}}\,{\lambda_{\mathrm{Nb}}}\,{c_{\mathrm{Nb}}}{(1\,-\,{f_{\mathrm{met.}}})}]} {\sum_{i} A_{i}~/~{({\delta_i}\,{\lambda_i}\,{c_i})}}}\,\times\,N_{\mathrm{t}}. 
\end{eqnarray}
In the case of a multicomponent system, this quantification is subject to several assumptions and simplifications. To specifically study the interaction between Nb and impurities, we first subtract the metallic Nb component ($f_\mathrm{met.}$) using the Nb 3d data. Due to the presence of multiple impurities, we estimate the total number of electrons involved in bonding ($N_\mathrm{t}$) and the number of bonds by using a weighted average based on impurity concentrations ($c_\mathrm{i}$). Since the homogeneity of the system is uncertain, we utilize matrix factor approximations. For carbon and oxygen, we assume the hybridization of 2s and 2p orbitals. We calculate the Gaussian peak area ($A_\mathrm{i}$) and obtain the values for cross-section area ($\sigma_\mathrm{i}$) and inelastic mean free path ($\lambda_\mathrm{i}$) from literature \cite{SunRef68,SunRef69}. Lastly, molecular orbital theory is not involved. By following these methods, we can effectively compare samples processed by different heat treatments (although the absolute values should be refined). For instance, in figure~\ref{SunFig7}, the electron populations observed at different depths after the 800\textdegree C UHV baking, regardless of the heating history, exhibit striking similarities, indicating the intrinsic effect on bonding of this heat treatment. 

Figure~\ref{SunFig7}(a) demonstrates the versatile capability of suboxides to arrange the number of electrons in the presence of carbides. The 120\textdegree C low-temperature baking reveals the effect of air exposure on the carbide-populated surface, showing that oxygen captures more electrons and creates more ionic character, especially within the first 10\,nm. In figure~\ref{SunFig7}(b), the air exposure of the N$_2$-processed sample and its EP'ed sample exhibit similar behavior, with a more significant impact that results in nearly no electrons around the Nb orbital within the first 5\,nm. The topic of ionic bonding is relevant to theories for high $T_\mathrm{c}$ oxides and carbides \cite{SunRef70}, but caution should be taken when interpreting it in the Nb system, where BCS theories apply.

Moreover, mid-temperature baking at 300\textdegree C generates a uniformly distributed region of high-electron-affinity oxygen, coinciding with the decomposition of higher-order oxides on the surface. However, this effect gradually diminishes as the baking temperature increases to 400\textdegree C and 800 \textdegree C, yielding a slope of oxygen's electron population as a function depth. In contrast, the N$_2$-processed samples at 800\textdegree C, as well as those UHV-baked at 120\textdegree C and 300\textdegree C, induce more ionic character at the surface region or exhibit a uniform profile of high ionic character throughout the sample. Overall, the carbide-facilitated oxygen diffusion and surface higher-order oxide decomposition affect the electron distribution in the suboxides mixed within the superconducting Nb.

\subsection{Nb coupon studies}
\begin{figure*}[b!]
\centering
\includegraphics[width= \linewidth]{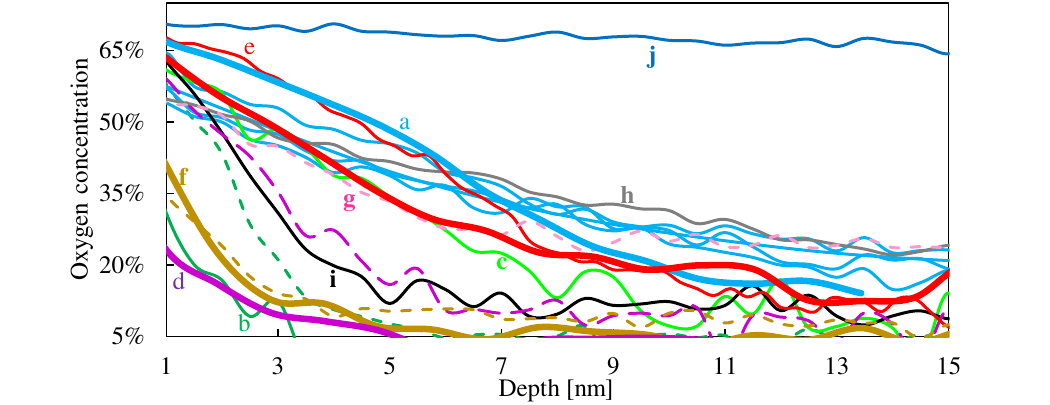}
\caption{Coupon composition analysis: Oxygen concentration as a function of depth on Nb coupons measured with the SSX-100 XPS, compared with thicker lines representing \textit{in situ} data taken with the PHI XPS. (a) EP'ed (light blue), (b) BCP'ed (dark green), (c) HF soaked (light green), (d) 800\textdegree C baked (purple), (e) "N$_2$ doped" and exposed to air (red), (f) "N$_2$ doped" and EP'ed (orange), (g) "N$_2$ infused" (pink), (h) modulated 20\%\,O$_2$\,/\,80\%\,N$_2$ flow for $>$\,1 day (grey), (i) ozone treated for 3 days (black), and (j) electrochemically anodized (dark blue). The dashed lines indicate samples exposed to air for $>$\,6 months.}
\label{SunFig10}
\end{figure*}

\begin{figure*}[b!]
\centering
\includegraphics[width= \linewidth]{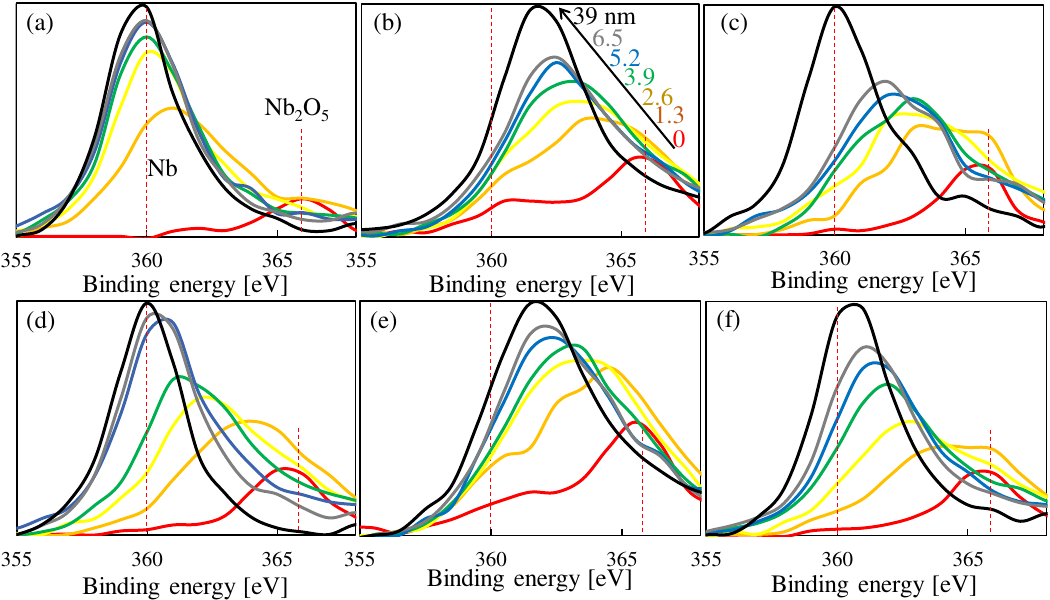}
\caption{Coupon spectral analysis: Depth-dependent XPS 3p$_{3/2}$ spectra on Nb coupons measured at depths of 0\,--\,6.5\,nm and 39\,nm. Acid/polishing treatments: (a) BCP'ed, (b) EP'ed, and (c) HF soaked. These samples were stored in air for $<$\,3 days before measurement. Effects of oxidizer exposure: (d) BCP'ed sample after $>$\,6 months in air, (e) EP'ed sample exposed to a modulated 20\%\,O$_2$\,/\,80\%\,N$_2$ flow for $>$\,1 day, (f) EP'ed sample exposed to ozone for 3 days. The binding energies of different oxide structures in the Nb 3p spectra are indicated in figure~S13. The intensity units are arbitrary.}
\label{SunFig11}
\end{figure*}

Qualitative analyses are conducted on Nb coupons listed in table~\ref{tbl:2} using Nb 3p and O 1s data collected with the SSX-100 XPS instrument. The concentration values obtained from high-intensity survey scans are reliable, while the Nb 3p spectra, which contain well-separated spin orbitals and better calibration in the SSX-100 XPS instrument, are used for comparison.  

The oxygen concentration profiles in figure~\ref{SunFig10} demonstrate that the Nb surface is highly sensitive to various acid-related and heat treatments, as well as environmental factors. The EP'ed samples (light blue) have been used as a calibration reference between different samples. We find a significantly higher quantity of higher-order oxides with greater depth in the EP'ed and N$_2$-processed samples, regardless of whether they are N$_2$ "doped" (red) or "infused" (pink), which is consistent with our \textit{in situ} investigations. In contrast, the BCP'ed (green), 800\textdegree C UHV baked (purple), and "N$_2$ doped"\,$+$\,EP'ed (orange) samples exhibit a thinner higher-order oxide layer on the surface. 

After 6 months of exposure to ambient air, most samples develop a typical oxide stack that matches a reference oxide model derived from ozone treatment (black). However, the "N$_2$-doped"\,$+$\,EP'ed sample (orange) does not show an increase in higher-order oxide thickness after the same exposure period. This suggests that passivating oxides are generated under this unique condition, and the EP mechanism for this type of sample requires further investigation. 

Furthermore, figure~\ref{SunFig11} presents the oxide structural differences imposed by acid/polishing and oxidizer exposure factors, showing similar observations.

\begin{figure*}[b!]
\centering
\includegraphics[width= \linewidth]{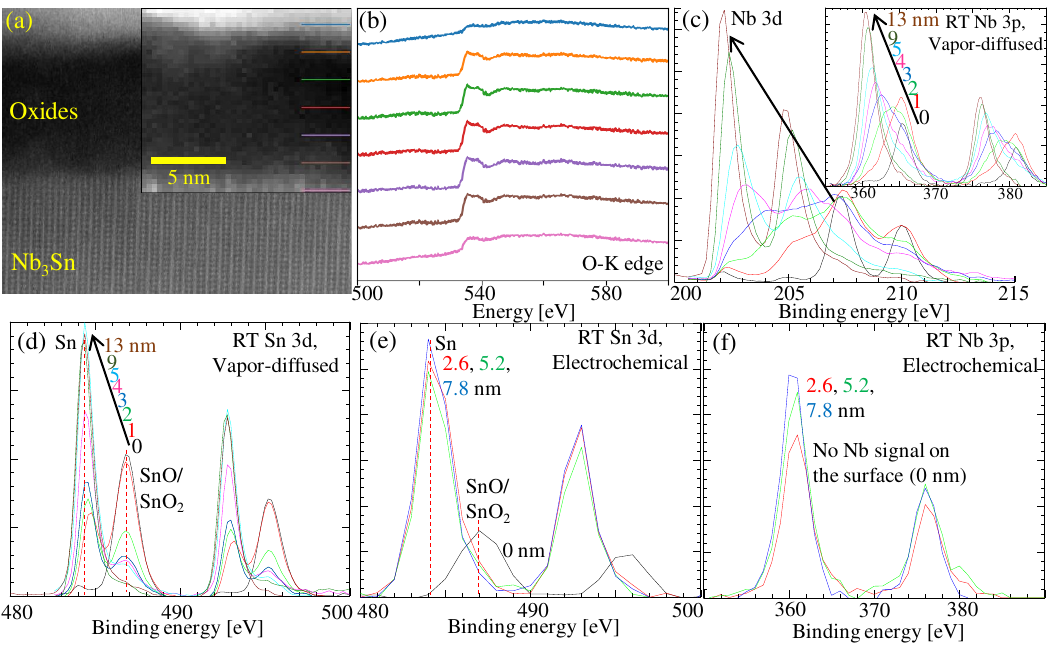}
\caption{Native oxides on Nb$_3$Sn at RT: (a) Cross-sectional STEM image showing surface oxides on Nb$_3$Sn produced by vapor diffusion. (b) EELS O-K edge spectra, collected at various depths as indicated in (a). (c) Nb 3d and 3p spectra and (d) Sn 3d spectra as a function of depth, measured with the PHI XPS on vapor-diffused Nb$_3$Sn. (Data measured with the SSX-100 XPS is included in figure~S14). (e) Sn 3d and (f) Nb 3p spectra as a function of depth measured with the SSX-100 on electrochemical Nb$_3$Sn. The intensity units in (b\,--\,f) are arbitrary.}
\label{SunFig12}
\end{figure*}
\subsection{Native oxides on Nb$_3$Sn at RT}
Characterizing surface oxides on Nb$_3$Sn is crucial due to the current challenges faced by Nb$_3$Sn cavities, which are hindered by unclear quench mechanisms \cite{SunRef48,SunRef49}. The theory surrounding how surface oxides influence Nb$_3$Sn cavities is evolving \cite{SunRef22,SunRef24}. Here we report our findings on oxides in both vapor-diffused and electrochemically synthesized Nb$_3$Sn, the only two types capable of producing high-performance SRF cavities to date. Our observations of distinct oxide structures and thicknesses provide atomic models for the development of emerging theories.

Figure~\ref{SunFig12}(a) displays a cross-sectional STEM image of vapor-diffused Nb$_3$Sn, revealing an amorphous oxide region with a thickness exceeding 9\,nm. EELS O-K edge spectra (figure~\ref{SunFig12}(b)), while slightly affected by the Sn-M edge, exhibit nearly identical features that indicate a mixed structure throughout the oxide region. XPS Nb 3d and 3p spectra in figure~\ref{SunFig12}(c) demonstrate a similar oxide evolution as a function of depth compared to Nb, but with an almost doubled thickness. Notably, these Nb oxides are further mixed with either SnO$_2$ or SnO, which have close binding energies, as shown in figure~\ref{SunFig12}(d). 

Particularly noteworthy are the strikingly different surface oxides present on the electrochemically synthesized Nb$_3$Sn. As shown in figures~\ref{SunFig12}(e) and (f), this type of Nb$_3$Sn only has a surface layer of $<$\,2.6\,nm SnO$_2$/SnO, without any higher-order Nb oxides. The existence of suboxides needs further investigation.  
\subsection{Effects of UHV baking temperature on Nb$_3$Sn}
\begin{figure*}[t!]
\centering
\includegraphics[width= \linewidth]{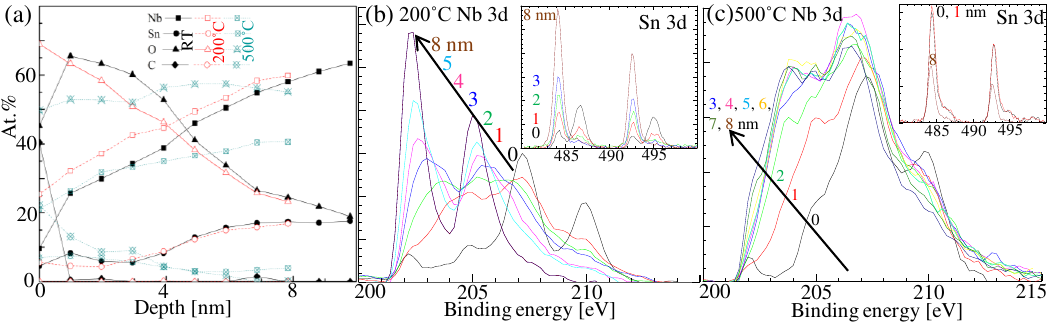}
\caption{UHV baking in Nb$_3$Sn oxides: (a) Concentration depth profiles of Nb, Sn, O, and C, collected by the PHI XPS on vapor-diffused Nb$_3$Sn at RT and during \textit{in situ} heating at 200\textdegree C and 500\textdegree C under UHV without exposure to air between measurements. (b,c) Nb 3d and Sn 3d spectra as a function of depth during \textit{in situ} heating at (b) 200\textdegree C and (c) 500\textdegree C. The intensity units in (b) and (c) are arbitrary.}
\label{SunFig13}
\end{figure*}

Upon \textit{in situ} heating at 200\textdegree C (figure~\ref{SunFig13}(b)), slight reconstruction and decomposition occur in both Nb oxides and Sn oxides mixed on the surface. However, these metastable oxides undergo dramatic decomposition and transformation during \textit{in situ} 500 \textdegree C baking. As shown in figure~\ref{SunFig13}(c), Sn oxides disappear, and Nb oxides begin to incorporate a significant quantity of oxides with all charges, resulting in wide spectra throughout the 8\,nm probed. Additionally, up to 20\,at.\% carbon appears on the surface. 

The valence spectra of these baking results in figure~S15 indicate the conducting behavior, most likely normal-conducting, of this thick oxide region.

\section{Conclusions}

In summary, we investigated the effects of different baking and processing conditions, including UHV baking, N$_2$ processing, acid/polishing, and oxidizer exposure treatments, on the structures, compositions, depths, and valence distributions of surface oxides, carbides, and impurities on Nb and Nb$_3$Sn.

Our findings revealed that these treatments significantly alter the type and thickness of Nb surface oxides. We found that the entire surface region (\textit{e.g.}, $\sim$\,7\,nm in EP'ed Nb), containing amorphous oxides and metallic components, is most likely normal-conducting, which can impact the superconducting bulk. Using XPS, STEM/EELS, and \textit{in situ} measurements we accurately established the structural profiles of the oxide regions as a function of depth under various baking and processing conditions.

Moreover, we resolved the underlying mechanism of baking and N$_2$-processing in Nb (which had previously achieved record SRF performance). We discovered that high-temperature baking induces carbide formation, regardless of gas environment. This, in turn, determines the subsequent formation of suboxides and higher-order oxides upon air exposure and low-temperature baking. These findings, combined with previous studies \cite{SunRef23,SunRef19,SunRef20,SunRef24}, strongly support that both surface oxides and second-phase formation collectively contribute to the effects induced by UHV baking (or oxygen processing) and nitrogen processing. Specifically, UHV baking at 500\textdegree C and N$_2$ processing at 800\textdegree C, with significant carbide formation, introduce a large number of suboxides and create more ionic character toward the surface after air exposure and 120\textdegree C\,--\,300\textdegree C baking. Additionally, the decomposition of surface higher-order oxides at 250\textdegree C\,--\,300\textdegree C also modifies the electron population profile at the surface.

Furthermore, our investigation revealed that vapor-diffused Nb$_3$Sn contains $>$9\,nm thick oxides mixed with Nb oxides and Sn oxides, while electrochemically synthesized Nb$_3$Sn only has $<$\,2.6\,nm thin Sn oxides on the surface. The thick metastable oxides on vapor-diffused Nb$_3$Sn may pose critical issues.

Overall, this study provides important insights into the fundamental mechanisms of baking and processing Nb and Nb$_3$Sn surface structures for high-performance SRF and quantum applications.  

\section{Supporting Information}
The Supporting Information is submitted.


\section{Data Availability Statement}
Data are available by contacting the corresponding author.

\section{Conflicts of Interest}
The authors declare no competing financial interest.

\section{Acknowledgement}
This work was supported by the U.S. National Science Foundation under Award PHY-1549132, the Center for Bright Beams. Utilization of the PHI Versaprobe III XPS within UVa's Nanoscale Materials Characterization Facility (NMCF) was fundamental to this project; we acknowledge NSF MRI award $\#$1626201 for the acquisition of this instrument. This work also made use of the Cornell Center for Materials Research Shared Facilities which are supported through the NSF MRSEC program (DMR-1719875), and was performed in part at the Cornell NanoScale Facility, an NNCI member supported by NSF Grant NNCI-2025233. Z.S. thanks H. G. Conklin and T. M. Gruber for assisting with sample preparation and Dr. M. Salim for XPS assistance. 

\section*{References}
\bibliographystyle{iopart-num}
\bibliography{ref}

\end{document}